\documentclass[11pt]{article}
\pdfoutput=1

\usepackage[letterpaper,
width=6.5in, 
left=1in,
top=1.3in,
bottom=1.15in,
headheight=19pt,
]{geometry}

\usepackage[utf8]{inputenc} 
\usepackage[T1]{fontenc}
\usepackage{url}            
\usepackage{booktabs}       
\usepackage{cite}
\usepackage{graphicx}
\usepackage{floatflt}
\usepackage{amssymb}
\usepackage{amsmath}
\usepackage{xfrac}
\usepackage{xspace}
\usepackage{enumitem}
\usepackage[usenames,dvipsnames]{color}
\usepackage{tikz}
\usepackage[normalem]{ulem}
\usepackage{wrapfig}
\usepackage[hang,flushmargin]{footmisc}

\usepackage{caption}
\DeclareCaptionFont{xipt}{\fontfamily{fvs}\fontsize{9}{11}\selectfont}
\usepackage[labelfont={xipt,bf},textfont={xipt},font={xipt}]{caption}

\usepackage{multirow}
\usepackage[square,sort,comma,numbers]{natbib}

\setlist{nolistsep} 
\definecolor{darkgreen}{RGB}{0,192,0}
\definecolor{steelblue}{RGB}{182,221,232}
\definecolor{lightorange}{RGB}{252,213,180}
\definecolor{darkorange}{RGB}{220,100,0}
\definecolor{lightmagenta}{RGB}{204,192,218}
\definecolor{gray}{RGB}{128,128,128}

\newcommand{\Architecture}{Flow\xspace}
\newcommand{\techterm}[1]{{\sffamily\selectfont{#1}}}

\graphicspath{{./graphics/}{./}}

\usepackage{ntheorem}
\theoremstyle{plain}
\newtheorem{theorem}{Theorem}

\usepackage{hyperref}       
\hypersetup{
	colorlinks=true,       
	linkcolor=[rgb]{0,0.4,0.5},          
	citecolor=[rgb]{0,0.3,0.3},        
	urlcolor=[rgb]{0,0.3,0}           
}

\title{\Architecture: Separating Consensus and Compute}
\date{} 

\begin{document}
\maketitle

\begin{center}
	Dr.\ Alexander Hentschel \\
	\footnotesize\texttt{alex.hentschel@dapperlabs.com} \\[10pt]
	\large
	\begin{tabular}{c@{\hspace{60pt}}c}
		Dieter Shirley & Layne Lafrance \\
		\footnotesize\texttt{dete@dapperlabs.com} & \footnotesize\texttt{layne@dapperlabs.com} 
	\end{tabular}
\end{center}
\vspace{10pt}

\begin{abstract}
\noindent
Throughput limitations of existing blockchain architectures are well documented and are one of the most significant hurdles for their wide-spread adoption. 
Attempts to address this challenge include layer-2 solutions, such as Bitcoin's Lightning or Ethereum's Plasma network, that move work off the main chain.
Another prominent technique is sharding, i.e.,\ breaking the network into many interconnected networks.
However, these scaling approaches significantly increase the complexity of the programming model by breaking ACID
guarantees (Atomicity, Consistency, Isolation, and Durability), increasing the cost and time for application development.

In this paper, we describe a novel approach where we split the work traditionally assigned to cryptocurrency miners into two different node roles.
Specifically, the selection and ordering of transactions are performed independently from their execution. 
The focus of this paper is to formalize the split of consensus and computation, and prove that this approach increases throughput without compromising security.

In contrast to most existing proposals, our approach achieves scaling via separation of concerns,
i.e.,\ better utilization of network resources, rather than sharding. 
This approach allows established programming paradigms for smart contracts (which generally assume transactional atomicity) to persist without introducing additional complexity.
We present simulations on a proof-of-concept network of 32 globally distributed nodes. 
While the consensus algorithm was identical in all simulations (a 2-step-commit protocol with rotating block proposer), 
block computation was either included in a consensus nodes' regular operations (conventional architecture) 
or delegated to specialized execution nodes (separation of concerns).
Separation of concerns enables our system to achieve a throughput increase by a \emph{factor} of 56 
compared to conventional architectures without loss of safety or decentralization.
\end{abstract}

\newpage
\tableofcontents

\newpage
\section{Introduction\label{sec:introduction}}   

The original Bitcoin blockchain uses a consensus model referred to as Nakamoto consensus \cite{nakamoto2012bitcoin}. 
It uses a sequential model in which a block is built, mined, 
and verified, and consensus about it is formed by nodes building subsequent blocks on top of it.
While the proof-of-work challenge (mining) that must be solved for every block provides tamper-resistance for the chain,
the associated computational effort limits block-production rate and transaction throughput. 
The throughput limitations of existing blockchain architectures are well documented and among the most 
significant hurdles for the wide-spread adoption of decentralized technology 
\cite{Kyle2016, BBC:CK_slows_Ethereum:2017, Richard_et_al:Bitcoin_Ethereum_Scalability:2019}.

The leading proposals for removing the overhead of proof-of-work adapt 
Byzantine Fault Tolerant (BFT) consensus algorithms \cite{Reaching_Agreement_in_Presence_of_Faults:Pease:1980, Lamport:1982:BFT}.
Blockchains using proof-of-work (PoW) require exceptional computational effort and, subsequently, electric power.
In contrast, finalizing blocks through BFT consensus is highly efficient but requires a known set of
participants, a super-majority of which must be honest.

Combining BFT consensus with proof-of-stake (PoS) \cite{Castro:1999:PBF:296806.296824} allows for the creation of 
a permissionless network with strong security properties.
Under PoS, all participating nodes are required to deposit a (financial) stake that can be taken away if they violate the protocol's rules. 
The amount of influence given to each node is proportional to its fraction of total stake.
Then the economic pressure (i.e.,\ the stake at risk) 
to follow the protocol is correlated with a node's  influence.
In addition, the deposited  stake has the added benefit of preventing Sybil attacks 
\cite{KingNadal:2012:PPCoin, AbrahamMalkhi:2017:BlockchainConsensusLayer, Gupta:2018:PWW}.
PoS systems promise to increase the throughput of the chain while also decreasing the total capital costs associated
with maintaining the security of the chain.
Even including the performance increases through adopting PoS, 
throughput restrictions are remaining the major challenge for wide-spread adoption
\cite{bano:2017:consensus, Spasovski:2017:PoS_Scalability}.%
\bigskip 

\noindent%
In this paper, we explore a novel approach to increasing the throughput of PoS blockchains.
While PoS blockchain proposals remove proof-of-work as the dominant sink of computational effort, 
they tend to inherit most of their architecture from the proof-of-work systems of the previous generation. 
In particular, every full node in the network is required to examine and execute each proposed block to update their local copy of the blockchain's state. 
As every transaction needs to be processed by every single node, adding nodes to the system provides no benefit in throughput or scale.  
Instead, adding nodes \emph{reduces} the throughput of most BFT consensus protocols, 
because the message complexity to finalize a block increases  
super-linearly with the number of consensus nodes (see section  \ref{sec:theo.perf.analysis:security-nodes}  for details).
Consequently, most PoS blockchains have to make a trade-off between a small consensus committee (weakening security)
or a low block production rate (decreasing throughput). 

For networks unwilling to compromise either security or decentralization, the most common approach to addressing the scaling
problem has been through sharding \cite{Kyle2016} or moving work off the main
chain (e.g.\ Lightning \cite{Bitcoin_Lightning:2016} or Plasma \cite{poon2017plasma} network). 
Both of these approaches, unfortunately, place significant limitations on the ability of transactions to access state
that is distributed throughout the network \cite{Zamani2018RapidChainAF}. 
These limitations dramatically increase the complexity required for developers who wish to deploy smart contract-based applications. 
\bigskip 

\noindent%
Our proposal, the \Architecture architecture, addresses these limitations by fundamentally changing how the blockchain is formed.  
\Architecture decouples the selection and ordering of transactions from their execution so that both processes can run in parallel.
The decoupling enables significantly higher transaction throughput than other blockchains architectures, without undermining security
or participation rates.

Traditional blockchain architectures require a commitment to the result of each block's state update to be included as part of the consensus process. 
As a result, every node must reproduce the state-update computation before it can finalize a block. 
Our finding is that consensus on the order\footnote{%
	In the full \Architecture architecture, Consensus Nodes work with transaction batches (\techterm{collections}).
	A block contains collection hashes and a source of randomness, which the Execution Nodes use to  shuffle the transactions before computing them. 
	While Consensus Nodes don't directly compute the transaction order for a block,
	they \emph{implicitly} determine the order by specifying all the inputs to the deterministic
	algorithm that computes the order.
	The detailed protocol is specified in a follow-up publication.
} 
of transactions in the block is all that is required.
Once that order is fixed, the resulting computation is determined even though it may not necessarily be known. 
Thereby, the computational effort of participating in consensus is significantly reduced, even for a very large number of transactions.
Once the transaction order is determined, ensuing processes can be delegated to perform the computation itself, without affecting decentralization of the system.

Our main body of work is section \ref{sec:TheoreticalAnalysis}, where we discuss and prove our central theorem,  
which states that one can separate the majority of computation and communication load from consensus without compromising security. 
Section \ref{sec:EmpiricalResults} complements the theoretical discussion by benchmarking the \Architecture architecture on an experimental 
network.
We conclude the paper by outlining the future work that would be required to implement a system based on these ideas in section \ref{sec:further_work}.
\medskip

\noindent%
The focus of this paper is to formalize the split of consensus and computation 
and prove that this approach increases throughput 
while maintaining strong security guarantees.  
A blockchain designed around the principles outlined in this paper would need to specify a variety of additional mechanisms including:
\begin{itemize}
	\item a full protocol for verifying computation results,
	\item details of the consensus algorithm,
	\item and adequate compensation and slashing mechanics to incentivize nodes to comply with the protocol.
\end{itemize}
Detailed formalization and analysis of these topics is reserved for follow-up publications. 

\subsection{Terminology\label{sec:Nomenclature}}

While most current blockchains focus solely on processing financial transactions,  
we consider a blockchain as a general, Turing-complete, distributed computing platform. Instead of referring to the blockchain as a ledger,  
we adopt the terminology of a distributed state machine wherein transactions describe transitions between computational states. 
Furthermore, we use the term \techterm{consensus} to refer only to linearizing the order of state transitions 
(but do not consider agreement about the computational result as a part of consensus).

\subsection{Related Work\label{sec:RelatedWork}}

\noindent
Blockchains supporting Turing-complete computation generally impose an upper limit on the computation within one block, such as Ethereum's \techterm{gas limit}.
Such a gas limit, in turn, introduces undesired throughput restrictions. 
One reason for imposing a gas limit in the first place is to avoid the Verifier's Dilemma \cite{Luu:2015:VerifierDilemma}. 
By setting the gas limit low enough, the time investment for verification is  negligible compared to solving the PoW challenge for mining the block. 
Thereby, the gas limit ensures that performing the verification work does not introduce a pivotal disadvantage for a node to successfully mine the next block. 

For PoS blockchain, the Verifier's Dilemma persists when incentives are given for speedy operation.  
Especially for high-throughput blockchains, the verification of a large number of transactions consumes significant computational resources and time. 
By separating consensus and computation, the Verifier's Dilemma for checking correctness of the execution result is of no concern anymore to the consensus nodes\footnote{%
	The Verifier's Dilemma (checking the correctness of the execution result) is of no concern anymore to the Consensus Nodes. 
	However, checking cryptographic signatures and proofs in PoS can still require noticeable computational work. 
	Delay through verifying signatures might induce a Verifier's Dilemma for the consensus nodes 
	Though, on a much smaller scale compared to requiring the consensus nodes to re-execute all transactions. 
}.
The maximum amount of computation in a block (the block's gas limit) can now be increased  
without the consensus nodes affected by  Verifier's Dilemma or slowed down by the computation work. 
However, the Verifier's Dilemma  still needs to be solved for the verifiers of the computation. 
Potential solutions include \techterm{zkSnarks} \cite{Bitansky:2012:zkSnark}, \techterm{Bulletproofs}  \cite{Bulletproofs:Buenz:2018},
and \techterm{TrueBit}'s approach \cite{teutsch:2017:Truebit_VerifierDilemma-Solution}.
For \Architecture, we developed \techterm{Specialized Proofs of Confidential Knowledge} (\techterm{SPoCK}s) to overcome the Verifier's Dilemma, 
which is described in detail in a follow-up paper.

Another limitation for blockchains are the resource demands to store and update the large computational state. 
In Ethereum, the gas limit is also used to control the rate at which the state grows \cite{Buterin:Comment_On_State_growth:2018}.
By delegating the responsibility to maintain the large state to specialized nodes, 
hardware requirements for consensus nodes can remain moderate even for high-throughput blockchains.
This design increases decentralization as it allows for high levels of participation in consensus 
by individuals with suitable consumer hardware on home internet connections. 
\medskip 

\noindent%
The concept of separating the issue of transaction ordering from the effort of computing the results of the computations has been previously
explored in the context of distributed databases \cite{FaunaDB:2016, AmazonAurora:2018}.   
Here, transactions are ordered into a log through a quorum (consensus), and subsequently each node can independently resolve transaction effects. 
However, these systems are designed for operation in well-maintained data-centers where Byzantine faults are not concern.
Specifically, the number of participating nodes is small, and node failure modes are restricted to dropouts. 

Within the blockchain space, the Ekiden  \cite{OasisEkiden:2018} paper describes a system where consensus is separated from computation,
with the goal of preserving the privacy of the contract execution environment.
The paper, which explains part of the Oasis blockchain technology \cite{Song:2018:Oasis_MS_presentation}, 
 notes that this approach leads to improved
performance. But it does not quantify the performance gain or prove that the resulting system maintains security.

\clearpage
\section{Architecture Overview\label{sec:ArchitectureOverview}}

The \Architecture architecture is founded on the principle of `separation of concerns'.
The network has specialized roles: 
\techterm{Consensus Nodes} and \techterm{Execution Nodes}.
The core differentiation between the two node types is 
\techterm{Objectivity} vs.\ \techterm{Subjectivity}.
\emph{Objective tasks} are those for which there is an objectively correct answer.
Any traditional mathematical calculation is objective; you don't need an authority or expert to
confirm the correctness of $2 + 2 = 4$, or to confirm that an actor that claims $2 + 2 = 5$ is Byzantine.
\emph{Subjective tasks} have no such deterministic solution. Most human governance systems (``laws'')
typically deal with subjective issues. At different times in different societies, the rules
about who can do certain things can be very different. 
The definition of the word \techterm{consensus} means the agreement on  subjective problems, where there is no single correct answer. Instead, one answer must be selected through mutual agreement. 

\begin{figure}[b!]
	\centering
	\vspace{-5pt}
	\includegraphics[width=0.7\textwidth, trim=35 275 35 170, clip]{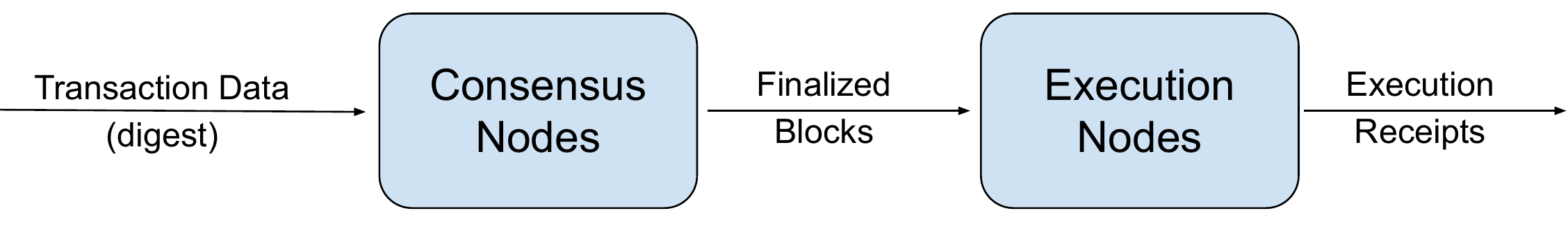}
	\caption{
		Overview of the message flow through Consensus and Execution Nodes.
		For brevity, only the messages during normal operation are shown.
		Messages that are exchanged during the adjudication of slashing requests are omitted. 
	}
	\label{fig:Role_Illustration}
	\vspace{-10pt}
\end{figure}

Blockchains combine objective rules with a decentralized mechanism
for resolving subjective problems. One example is if two transactions are submitted at the same
time that try to spend the same coins (e.g.,\ no double-spends), which one resolves correctly, and which one fails?
Traditional blockchain architectures ask the nodes participating in the network to solve both kinds of problems at the same time.
In \Architecture, the Consensus Nodes are tasked with all subjective questions,
while the Execution Nodes are responsible solely for fully deterministic, objective problems.
While we reserve a detailed discussion of the nodes' roles, tasks, and interactions for the follow-up paper,
we briefly define the \techterm{Consensus Role} and \techterm{Execution Role} for nodes.
For an illustration, see Figure \ref{fig:Role_Illustration}.

\subsubsection*{Consensus Role}
Consensus Nodes form blocks from transaction data digests. 
Essentially, Consensus Nodes maintain and extend the core \Architecture blockchain. 
An agreement to accept a proposed block needs to be reached by many nodes 
which requires a \techterm{Byzantine-Fault-Tolerant (BFT)} consensus algorithm
\cite{Reaching_Agreement_in_Presence_of_Faults:Pease:1980}.
It should be noted that the results in this paper hold for \emph{any BFT consensus algorithm} 
with deterministic finality. 

In \Architecture, a block references its transaction and defines their execution order. 
However, a block contains \emph{no} commitment to the resulting computational state \emph{after} block execution. 
Accordingly, Consensus Nodes do not need to maintain the computational state or execute transactions. 

Furthermore, Consensus Nodes adjudicate slashing requests from other nodes, for example
claims that an Execution Node has produced incorrect outputs.

\subsubsection*{Execution Role}

Execution Nodes provide the raw computational power needed to determine the result of the
transactions when executed in the order determined by the Consensus Nodes.
They produce cryptographic attestations declaring the result of their efforts in the form of \textsf{Execution Receipts}.
These receipts can be used to challenge the claims of an Execution Node when they are shown to be incorrect. 
Also they are used to create proofs of the current state of the blockchain once they are known to be correct.
The verification process -- by which Byzantine Receipts are rejected (and the Execution Nodes which produced them are slashed),
and by which valid receipts are accepted (and shared with observers of the network) -- is outside the scope of this paper and
will be addressed in future work.

\clearpage
\section{Theoretical Performance and Security Analysis\label{sec:TheoreticalAnalysis}}

\noindent%
In this section, we present a theoretical derivation that  one can separate 
the majority of computation and communication load from consensus nodes without compromising security. 
Furthermore, we provide an analysis that explains the source of the experimentally observed throughput increase. 

\Architecture is designed to guarantee\footnote{%
	In \Architecture, guarantees are probabilistic. Specifically, errors are detected and corrected with 
	probability $p = 1 - \varepsilon$ for $0 < \varepsilon \ll 1$. Through system parameters, 
	$\varepsilon$ is tuned such that the desired properties hold with near certainty. 
}
that any error introduced by the Execution Nodes maintains four critical attributes:
\begin{itemize}
	\item \textbf{Detectable}:
	A deterministic process has, by definition, an objectively correct output. 
	Therefore, even a single honest node in the network can detect deterministic faults, 
	and prove the error to all other honest nodes by pointing out the part of the process that was executed incorrectly.
	\item \textbf{Attributable}:
	The output of all deterministic processes in \Architecture must be signed with the identity of the node that generated those results. 
	As such, any error that has been detected can be clearly attributed to the node(s) that were responsible for that process.
	\item \textbf{Punishable}:
	All nodes participating in a \Architecture network, including Execution Nodes, 
	must put up a stake that can be slashed if they are found to have exhibited Byzantine behaviour. 
	Since all errors in deterministic processes are detectable and attributable, those errors can be reliably punished via slashing.
	\item \textbf{Recoverable}:
	The system must have a means to undo errors as they are detected. 
	The property serves to deter malicious actors from inducing errors that benefit them more than the slashing penalty.
\end{itemize}
\smallskip

\noindent%
An important property of this design is that for each system-internal operation, the participants are accountable.
Specifically, for all operations \emph{except for the Consensus Nodes}, the execution of each operation is
delegated to a subset of nodes, the \techterm{operation processors}. 
Verifying the outcome is assigned to a disjoint node set, the \techterm{operation verifiers}.
Informally, the protocol works as follows:

\begin{itemize}
  \item Both \techterm{operation processor} and \techterm{operation verifier} groups are chosen at random. 
          The selection of nodes uses a verifiable random function \cite{Micali:1999:VRFs}, such that the outcome is
            deterministic but resistant to hash grinding.
  \item The inclusion probability for a node in either group is proportional to its stake.
			This enforces that Byzantine actors must lock up a significant amount of stake in
			order to have a non-negligible probability of affecting the system.
			Specifically, this hardens the system against Sybil attacks \cite{Douceur2002Sybil}.
  \item The required amount of stake for both groups is set sufficiently high such that the
            probability of sampling \emph{only} Byzantine actors in both groups is sufficiently small.
  \item As long as at least one honest node is involved in either group, the honest node
            will detect and flag any error.
  \item If a potential error is flagged, the case is adjudicated by the Consensus Nodes,
           malicious nodes are slashed, and the operation's outcome is rejected if faulty.
\end{itemize}
The process above guarantees that malicious Execution Nodes are  slashed with near certainty. Furthermore, it is nearly impossible for the malicious actors to succeed with introducing an error. 

A question that might arise is why \Architecture has separate groups of \techterm{operation processors} and  \techterm{operation verifiers} 
instead of the \techterm{operation processors} verifying each other's results. 
We chose this separation of concern to address the \techterm{Verifier's Dilemma} \cite{Luu:2015:VerifierDilemma}.
Without a dedicated verifier role, there is a conflict of interest for an \techterm{operation processor} to compute the next block 
vs.\ verifying (re-computing) the last block's result. 
In \Architecture, this dilemma is alleviated by dedicated \techterm{operation verifiers} who are compensated solely for verification.  
The technical details of our \techterm{block verification} protocol are presented in the follow-up paper, 
including a solution to the `freeloader problem' (verifiers just approving any result without doing the actual computation) 
and `maliciously flagging' (verifiers challenging  correct results to congest the network).

The following theorem formalizes the security guarantees of the \Architecture architecture 
and proves that introducing an error into the system by publishing or approving faulty results is economically infeasible.

\begin{theorem}[Probabilistic security for delegation of work to small groups]\label{cor:delegation_of_work}
$~$\\
Introducing an error into the system 
by deliberately publishing or approving faulty results is economically infeasible,
if the following conditions hold for any operation, except for those from the Consensus Nodes.
\begin{enumerate}
	\item The operation is delegated to two sets of randomly selected nodes:
			  \begin{enumerate}
				\item  set of  \techterm{operation processors}: members of this group execute the operation and provide cryptographically secure commitments to their result 
				\item  set of  \techterm{operation verifiers}: members of this group verify the operation's result and provide cryptographically secure commitments to the result if they approve
			  \end{enumerate}
	\item Both groups can be relatively small as long as the probability of choosing \emph{only} Byzantine actors in \emph{both} groups at the same time is sufficiently low.
	\item At the time the \techterm{operation processors} generate the result, the membership of the \techterm{operation verifier} group is unknown to them. 
    \item Consensus Nodes
			\begin{enumerate}
				\item  either verify that a significant majority have committed to the published outcome and there are no objections raised by participating nodes
				\item  or adjudicate objections, determine the faulty nodes (attributable), and slash them (punishable).
			\end{enumerate}          
\end{enumerate}
\end{theorem}

\noindent 
It is essential to highlight that Consensus Nodes are not required to check the
correctness of the results of an operation. Instead, they ensure that other nodes with
sufficient stake are accountable for the verification.
Furthermore, Theorem \ref{cor:delegation_of_work} holds for \emph{any} BFT consensus algorithm with deterministic finality.
\medskip

\noindent
\textbf{Proof of Theorem \ref{cor:delegation_of_work}:}
\\
We will show that Theorem \ref{cor:delegation_of_work} can always be satisfied under realistic conditions where
\begin{itemize}
	\item at least one of the two groups is sampled from a large population with a super-majority of honest nodes;
	\item Byzantine actors cannot suppress communication between correct nodes, 
	i.e.,\ if there is one honest node objecting to the result and proving its faultiness, 
	the erroneous nodes will be slashed.
\end{itemize}

\newpage
Specifically, let us consider a population of $N$ nodes from which we want to randomly draw a subset with $n$ nodes. 
Furthermore, we assume that there are at most $M < N/3$ Byzantine nodes. 
In the following, we focus on the case where all nodes are equally-staked%
\footnote{%
	The argument can be extended to nodes with different stakes. In this case, each node would 
	have an inclusion probability equal to its fraction of total stake. However, the probability 
	of sampling fully Byzantine groups is depending on the specific fractions of total stake 
	for the individual nodes.
	A basic solution for allowing nodes with different stakes is to introduce a unit quantity $\varrho$ of stake. 
	For a node with the stake $s$, the multiplicity $k = \lfloor s / \varrho  \rfloor$ represents how many \emph{full}
	staking units the node possesses. For operational purposes (including voting and node selection), 
	the blockchain treats the node identically to
	$k$ independent nodes each having stake $\varrho$. 
}, i.e.,\ their inclusion probabilities are identical. 
Drawing an $n$-element subset falls in the domain of simple random sampling \cite{olken1993random} without replacement.  
The probability of drawing $m \leq n$ Byzantine nodes is given by the hypergeometric distribution%
\footnote{%
   For conciseness, we only handle the case $m \leq M$. For $m > M$,  $\mathcal{P}_{_{n,N,M}}(m)  = 0$ per definition.
}
\begin{align}
  \mathcal{P}_{_{n,N,M}}(m) = \frac{{M \choose m} {N-M \choose n-m}}{{N \choose n}} \, .
\end{align}
The probability of a successful attack, $P(\textnormal{\textsf{successful attack}})$, 
requires that there is no honest node that would contradict a faulty result. Hence, 
\begin{align}\label{eq:delegation_of_work:success_prob_bound}
P(\textnormal{\textsf{successful attack}}) \leq \mathcal{P}_{_{n,N,M}}(n),
\end{align}
where $\mathcal{P}_{_{n,N,M}}(n)$ is the probability of sampling only Byzantine nodes.
\begin{align}
\mathcal{P}_{_{k,N,M}}(k) & = \frac{M!}{N!} \frac{(N-k)!}{(M-k)!} \qquad\textnormal{ for } k \leq M
\\\label{eq:delegation_of_work:proof_monotonous_decreasing_success_prob}
\Rightarrow\quad \frac{\mathcal{P}_{_{n,N,M}}(n)}{\mathcal{P}_{_{n+1,N,M}}(n+1)} &= \frac{(N-n)!}{(N-(n+1))!} \frac{(M-(n+1))!}{(M-n)!} = \frac{N-n}{M-n} > 1
\end{align}
As eq.\ \eqref{eq:delegation_of_work:proof_monotonous_decreasing_success_prob} shows,
the probability of sampling only Byzantine nodes is strictly monotonously decreasing with increasing $n$.
Eq.\ \eqref{eq:delegation_of_work:proof_monotonous_decreasing_success_prob} states that the larger the sample size $n$, 
the smaller the probability to sample only Byzantine node.

For a node to deliberately attack the network by publishing a faulty result or approving such, 
we assume the existence of some reward $r$ which the node receives in case its attack succeeds. 
However, if the attack is discovered, the node is slashed by an amount $\xi$ (by convention positive). 
The resulting statistically expected \textsf{revenue} from attacking the network is
\begin{align}
   \textnormal{\textsf{revenue}} &= P(\textnormal{\textsf{successful attack}}) \cdot r -  \big(1-P(\textnormal{\textsf{successful attack}})\big) \cdot \xi
   \\
   & \stackrel{\eqref{eq:delegation_of_work:success_prob_bound}}{\leq}
    \mathcal{P}_{_{n,N,M}}(n) \cdot r - (1-\mathcal{P}_{_{n,N,M}}(n)) \cdot \xi
\end{align}
For the attack to be economically viable, one requires $0 \stackrel{!}{\leq} \textnormal{\textsf{revenue}}$, which yields the central result of this proof: 
\begin{align}\label{eq:delegation_of_work:gain ratio}
\frac{r}{\xi} \ge \frac{1}{\mathcal{P}_{_{n,N,M}}(n)} - 1\,.
\end{align}
Furthermore, eq.\ \eqref{eq:delegation_of_work:proof_monotonous_decreasing_success_prob} 
implies that  $r/\xi$ increases strictly monotonously with increasing $n$. 
Using results from \cite{Hoeffding:1963}, one can show that $r/\xi$ grows exponentially with $n$ for $n < N/2$.

The left-hand side of equation \eqref{eq:delegation_of_work:gain ratio}, $r/\xi$, 
is a measure of  security as it represents the statistical \techterm{cost to attack} the system 
in the scenario where an attacker bribes nodes into byzantine behavior. 
As an example, let us consider the case with $N=1000$, $M=333$, $n = 10$.
For simplicity, assume that for publishing or approving a faulty result, the node's entire stake is slashed. 
Then, for an attack to be economically viable
the success reward $r$ would need to be $65\hspace{2pt}343$ times the node's stake.
If  the  \techterm{operation verifiers} staked \$\hspace{0.2pt}1000 each, 
an attacker would have to expend \$\hspace{1pt}65.3 million on average to cover all the slashing costs.
It would be cheaper for the attacker to run the entire pool of \techterm{operation verifiers} instead of attempting 
to slip an error past the honest verifiers. 
When increasing $n$ even further to  $n = 20$, an attacker would need to expend 
$r/\xi = 5.2 \cdot 10^9$ times the stake to slip a single error past a super-majority of honest verifiers. 

In summary, we have analyzed the case where either processing an operation or verifying its result
is delegated to a small, random subset of a much larger population of nodes. 
We have shown that under realistic assumptions, introducing an error into the system 
by publishing or approving faulty results is economically infeasible. 
Note that this result only covers node types \emph{other} than Consensus Nodes. 
Hence, it is sufficient for the Consensus Nodes to check that enough nodes have participated in executing the 
operation as well as verifying it. However, they do not need to check the result itself to problematically guarantee its integrity.
\vspace{-5pt}
\begin{flushright}
$\Box$
\end{flushright}

\noindent%
Theorem \ref{cor:delegation_of_work} is a key insight, as it allows us to:
\begin{itemize}
	\item separate the majority of computation and communication load from consensus;
	\item develop highly specialized nodes with distinct requirement profiles 
	        (as opposed to having one node type that has to have outstanding performance in all domains or otherwise diminish the network throughput);
\end{itemize}
While other nodes verify each others' operations in small groups, the entire committee of Consensus Nodes audits themselves.

\subsection{Special Role of Consensus Nodes\label{sec:theo.perf.analysis:security-nodes}}

Consensus Nodes determine the relative time order of events through a BFT consensus algorithm. 
While our results hold for \emph{any} BFT consensus algorithm with deterministic finality,
\techterm{HotStuff} \cite{HotStuff:2018, HotStuff:2019:ACM} is the leading contender. 
However, we continue to assess other algorithms such as 
\techterm{Casper CBC} \cite{Zamfir:CasperCBC_Template:2017, Zamfir:CasperTFG:2017, Zamfir_et_al:MinimalCasperFamily:2018}
or \techterm{Fant\protect\^omette} \cite{Azouvi:2018:Fantomette}.

In contrast to the operations of other nodes, which requires an auditing group to approve the result, 
Consensus Nodes finalize blocks without external verification. While the contents of a block can
be verified and Consensus Nodes punished if they include invalid entries, blocks are not rebuilt
in this scenario, unlike other verification processes.
External parties can inspect the 
finalized blocks after the fact.
However, in the event of a adversarial attack forking the chain,
a double-spend attack might have already succeeded at this point. 
To increase the resilience of the entire system, the committee of Consensus Nodes should consist of
as many staked nodes as possible. 

For a simple BFT algorithm, the message complexity $\eta$ per block (i.e.,\ the total number of messages sent by all $N$ nodes) is $O(N^2)$ \cite{Castro:1999:PBF:296806.296824}.
More advanced protocols achieve $\eta \in O(N\log N)$ \cite{Liu2019ScalableBC}
or $\eta \in O(c\cdot N)$ \cite{DBLP:journals/corr/abs-1804-01626, Proteus:Jalalzai:2019}, for $c \ll N$ an approximately constant value for large $N$. 
The overall bandwidth load $\mathcal{B}$ [$\textrm{MB}/\textrm{s}$] for the entire consensus committee is 
\begin{align}\label{eq:bandwidth_load}
  \mathcal{B} =   \beta \cdot b \cdot  \eta, 
\end{align} 
for $\beta$ the bock rate [$\textrm{s}^{-1}$], $b$ the message size [$\textrm{MB}/\textrm{s}$].  
For a node receiving a message $m$ and processing it, imposes a latency
\begin{align}\label{eq:message_latency}
\mathcal{L}(m) =   f(b) +\textrm{\texttt{process}}(m), 
\end{align} 
where $f$ denotes the network transmission time for receiving $m$ and $\textrm{\texttt{process}}(m)$ represents the computation time for processing $m$.
It is apparent that both $\mathcal{B}$ and $\mathcal{L}$ strongly impact the throughput of the consensus algorithm. 
The specific details are highly dependent on the chosen BFT protocol, as well as the employed gossip topology \cite{RandomizedGossipAlgorithms:Boyd:2006}.  
Other factors include delays or bandwidth limitations in the underlying network hardware.

Currently, we are targeting a consensus committee on the order of several thousand nodes. 
To support decentralization and transparency, hardware requirements for Consensus Nodes 
should be limited such that private groups can still afford to run a node and participate in consensus.   
Hence, given a desired minimal block rate (e.g.,\ $ \beta = 1\frac{\textrm{block}}{\textrm{s}}$) and an environment-determined function $f$,
the consensus committee can be increased only by decreasing message size $b$ or $\textrm{\texttt{process}}(m)$. 

For completeness, we provide a brief outlook on how we simultaneously reduce $b$ and $\textrm{\texttt{process}}(m)$ in \Architecture.
For the detailed mechanics of the respective processes, the reader is referred to subsequent papers. 
\begin{itemize}
	\item
	We delegate the computation of the transactions to the specialized Execution Nodes. 
	The delegation removes the need for Consensus Nodes to maintain and compute state, which significantly reduces  $\textrm{\texttt{process}}(m)$.
	\item
	Consensus Nodes do not require the full transaction texts during normal operation. 
	Instead, specialized \techterm{Collector Nodes} receive transactions from external	clients and prepackage them into batches, called \techterm{collections}.
	Consensus Nodes only order collection references (hashes),
	which substantially reduces the messages size $b$ compared to handing individual transaction texts.
\end{itemize}
\bigskip

\begin{wrapfigure}{r}{0.50\textwidth}
	\centering
	\vspace{-25pt}
	\includegraphics[width=0.5\textwidth, trim={0.8cm 7cm 0 6cm}, clip]{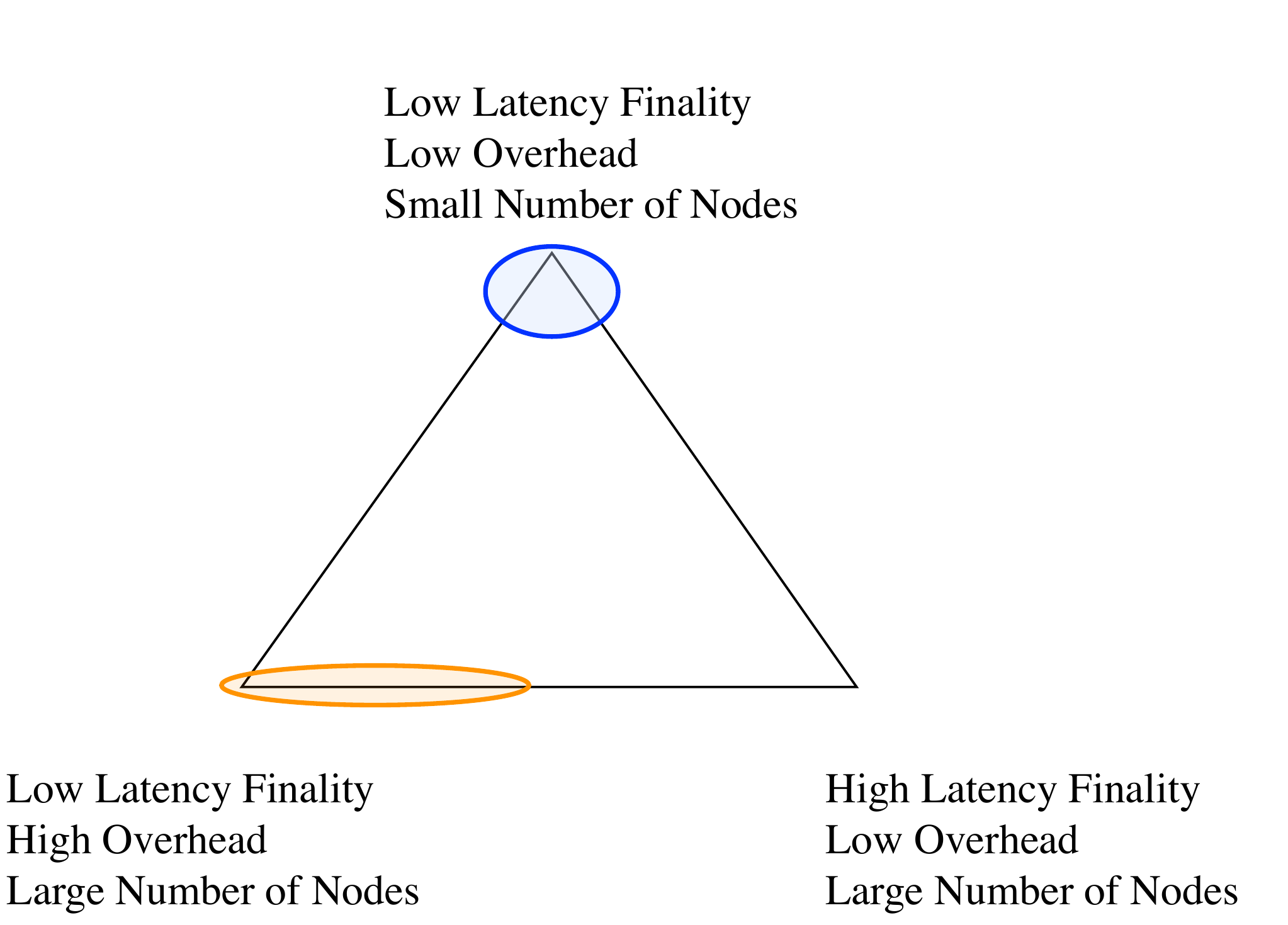}
	\caption{
		Zamfir's triangle of compromises; \\
		\textcolor{darkorange}{orange}: Consensus Nodes; \textcolor{blue}{blue}: Execution Nodes.
	}
	\label{fig:vlad_triangle_CN_AN_highlighted}
\end{wrapfigure}

\noindent
We conclude this section by comparing our approach with the `triangle of compromises'
proposed by Vlad Zamfir  \cite{Zamfir2017Triangle}, which we re-create in Figure \ref{fig:vlad_triangle_CN_AN_highlighted}. 
The triangle illustrates Zamfir's impossibility conjecture for proof of stake
systems.
While the conjecture has not been proven formally, we concur that the compromises are correctly identified. 
However, \Architecture optimizes where these compromises are made:

\begin{itemize}
	\item \emph{Consensus Nodes} work as part of a large consensus committee for maximal security. 
			To ensure security and fast generation of finalized blocks,
			we accept a higher communication overhead (the bottom-left corner of the triangle). However,
			unlike other blockchains, this consensus only determines the order of transactions within a block, but not the resulting state. 
			The architecture compensates for the resulting bandwidth higher overhead by minimizing
			message size through the use of collection references (references to batches of transactions) instead of full individual transactions.
			To further increase throughput and reduce communication latency, all possible computation 
			is delegated to other node types. 
\end{itemize}

\begin{itemize}			
	\item The \emph{Execution Nodes} unpack the collections and process individual transactions. 
			Without a large number of Byzantine actors involved, this can be handled directly between the Collector and
			Execution Nodes without involving Consensus Nodes. Furthermore, tasks can be parallelized and processed by small groups 
			which hold each other accountable for correct results. Only during Byzantine attacks,
			malicious actions would be reported to Consensus Nodes to adjudicate slashing. 
\end{itemize}

\clearpage
\section{Performance Simulations\label{sec:EmpiricalResults}}

The  theoretical analysis presented in section \ref{sec:theo.perf.analysis:security-nodes} suggests 
that transaction throughput can be increased by separating consensus about the transaction order from their execution. 
However, the theoretical analysis makes no assertion as to what the realistically achievable speedup is, 
as throughput heavily depends on a variety of environmental parameters  such as message round-trip time, CPU performance, etc. 
Therefore, we have implemented a simplified benchmark network that solely focuses on transaction ordering (consensus) and transaction execution. 

\subsection{Experimental Setup}
In a 2015 study analyzing the distribution of computational power in the Bitcoin network  \cite{Miller:2015:BitcoinTopology},
the authors estimated that 75\% of the mining power was provided by roughly 2\% of the nodes.
We simulated a system whose centralization metrics are roughly half of the Bitcoin scores.
In our simulations, roughly 38\% computational power is provided by the \textsf{fast nodes},
which represent approximately 6\% of the nodes.
For the remaining 62\% of the network's total computational power, we have applied a less-extreme ratio:
two-third of the nodes (\textsf{slow nodes}) hold one-third of the remaining computational power
(i.e.,\ $62\% / 3 \simeq 20\%$ of the total computational power).
The remainder is assigned to \textsf{medium nodes}.

To assign the Execution role to nodes with the most computation power 
requires incentive mechanisms  that compensate nodes for the resources used by the network. 
Assuming the existence of such incentive mechanisms, it is  economically rational for a \textsf{fast node} to stake as an Execution Node.
In any other role, its resources would not be utilized to the maximum potential leading to diminished revenue. 
Hence, we assumed that the most powerful nodes would stake specifically to become Execution Nodes. 
\medskip

\begin{figure}[t!]
	\centering
	\vspace{-10pt}
	\includegraphics[width=0.6\textwidth]{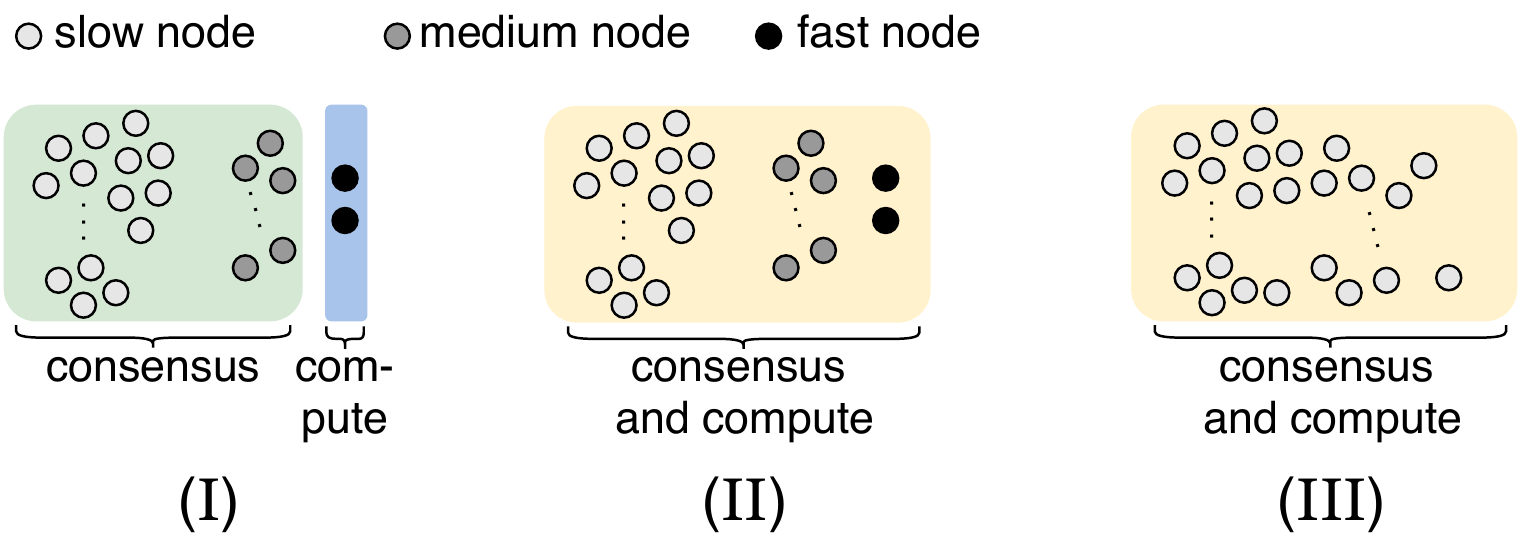}
	\caption{\textbf{Illustration of experiments.}
		\textbf{\textrm{(I)}} \Architecture network with the division of consensus and compute;
		\textbf{\textrm{(II)}} Conventional PoS network containing nodes with different performance levels;
		\textbf{\textrm{(III)}} Conventional PoS network containing only slow nodes.		
	}
	\label{fig:experiment_setup_vizualization}
	\vspace{-10pt}
\end{figure}

\noindent
We conducted three different experiments. The common characteristics of all simulations are described in the following. 
Section \ref{sec:ExperimentDescription:Bamboo} to \ref{sec:ExperimentDescription:Uniform-PoS} 
present the specific details for each individual experiment. 
Figure \ref{fig:experiment_setup_vizualization} illustrates the different setups. 
For each experiment, the network resources (node types) and the assignment of responsibilities 
are given in Table \ref{tab:experiment_setups:node_configuration}.

\begin{table}[!b]  
	\centering
	\begin{tabular}{lr|c|c|c}
		\toprule
		& & Slow Nodes & Medium Nodes & Fast Nodes \\ \midrule
		\textbf{Experiment (I)}: &  number of nodes &  20  & 10   & 2  \\
		& role of nodes & consensus  &  consensus  &  compute \\  \midrule
		\textbf{Experiment (II)} &  number of nodes &  20  & 10   & 2  \\
		& \multirow{2}{*}{role of nodes}  & consensus  &  consensus  &  consensus \\ [-3pt]
		&  & and compute  &  and compute  &  and compute \\  \midrule		
		\textbf{Experiment (III)} &  number of nodes &  32  &   &   \\
		& \multirow{2}{*}{role of nodes}  & consensus  &    &   \\[-3pt]
		&  & and compute  & & \\
		\toprule
	\end{tabular}
	\caption{Network configuration for each experiment. }
	\label{tab:experiment_setups:node_configuration}
\end{table}

\subsubsection*{Common Characteristics}

\emph{Transactions:} 
For simplicity, our network was processing benchmark transactions, which all had identical complexity (number of instructions). 
Batches of benchmark transactions were directly generated by the Consensus Nodes 
instead of integrating a dedicated submission process into the simulation. 
\smallskip

\noindent
\emph{Network:}
We implemented a relatively small network of 32 nodes, were each node ran on a dedicated Google Cloud instance. 
The nodes were spread over 8 data centres across the world to provide somewhat realistic latencies.
\smallskip

\noindent
\emph{Nodes:}
Depending on the experiment (see Table \ref{tab:experiment_setups:node_configuration}), 
transactions were executed on nodes with different hardware:
\begin{itemize}
	\item \textsf{slow nodes}: process a benchmark transaction in approximately 10ms
	\item \textsf{medium nodes}: five times as fast as slow nodes, i.e.,\ process  a benchmark transaction in 5ms
	\item \textsf{fast nodes}: 25 times as fast as slow nodes, i.e.,\ process a benchmark transaction in $2.5$ms
\end{itemize}
To facilitate decentralization, an ideal network should allow any participant to join, requiring only a minimal node performance. 
Consequently, a realistic network will contain a majority of slow nodes,  some medium nodes and very few fast nodes. 
\smallskip

\noindent
\emph{Consensus:}
We implemented a Tendermint-inspired consensus algorithm with a rotating bock proposer. 
As our goal was to benchmark achievable throughput in the absence of a large-scale Byzantine attack, 
our benchmark network only consists of honest nodes.
The proposed blocks contain a variable number of $t$ benchmark transactions, 
where $t$ is drawn uniform randomly from the integer interval $[240, 480]$.
However, for repeatability, we seeded the random number generator such that 
in all experiment, the same sequence of 20 blocks was proposed and finalized. 

\subsubsection{Experiment (I):  \Architecture PoS network with the split of consensus and compute\label{sec:ExperimentDescription:Bamboo}}
This experiment simulates a network with the division of consensus and compute:
\begin{itemize}
	\item 20 slow nodes and 10 medium nodes form the consensus committee. They agree on the order of transactions within a block but don't store or update the chain's computational state. 
	\item Two fast nodes execute the transactions in the order they are finalized in the blocks. They do not participate in consensus.
\end{itemize}
For further illustration, see Figure \hyperref[fig:experiment_setup_vizualization]{\ref{fig:experiment_setup_vizualization}(I)}
and Table \ref{tab:experiment_setups:node_configuration}.

In the \Architecture network, blocks only specify 
the transaction order, but there is no information about the resulting state included. As consensus in this model only covers the transaction order, consensus nodes are oblivious about the computational state. 

\subsubsection{Experiment (II):  PoS network of nodes with diverse performances\label{sec:ExperimentDescription:Conventional-PoS}}
Experiment (II) closely models conventional BFT consensus systems such as 
Tendermint \cite{Kwon2014TendermintC,  buchman2016tendermint, KwonAndBuchman:CosmosWhitepaper:2016}
or Hot-Stuff \cite{HotStuff:2018}.
The network is identical to \hyperref[sec:ExperimentDescription:Bamboo]{Experiment (I)}, i.e., it consists of exactly the same types and numbers of nodes 
(see Table \ref{tab:experiment_setups:node_configuration} and 
Fig.\ \hyperref[fig:experiment_setup_vizualization]{\ref{fig:experiment_setup_vizualization}(II)} for details and illustration). 
Though, blocks contain the transaction order 
and a hash commitment to the resulting computational state. 
Due to this result commitment, each consensus node must repeat the computation of all transactions for a proposed block to validate its correctness. 
In essence, there is only one role for a node: to participate in the consensus algorithm, which includes the task of updating the computational state.

\subsubsection{Experiment (III): PoS network of nodes with uniform performance\label{sec:ExperimentDescription:Uniform-PoS}}
Experiment (III), 
illustrated in Figure  \hyperref[fig:experiment_setup_vizualization]{\ref{fig:experiment_setup_vizualization}(III)},
simulates a network of 32 nodes with uniform computational performance.
As in \hyperref[sec:ExperimentDescription:Conventional-PoS]{Experiment (II)}, 
all nodes execute the same algorithm which combines consensus about transaction ordering with their computation. 

\subsection{Experimental Results}

Our simulations aim at benchmarking the transaction throughput. For each experiment, 
we sent 7995 benchmark transactions through the network 
and measured the corresponding processing time. 
The results are summarized in Table \ref{tab:empirical_results}. 

\begin{table}[b!]
	\begin{center}
		\begin{tabular}{r|cc}
			\toprule
			& \textbf{Processing Time [s]} & \textbf{Throughput [$\sfrac{\textrm{TX}}{\textrm{s}}$]} \\ \midrule
			Experiment (I)  & 5.14  & 1555.4          \\
			Experiment (II)  & 291  & 27.5            \\
			Experiment (III) & 293 & 27.3            \\
			\toprule
		\end{tabular}
		\caption{
			\textbf{Network performances}. Processing time for 7995 transactions in seconds [s] and the resulting transaction throughput [$\sfrac{\textrm{TX}}{\textrm{s}}$].
			While we only conducted one-shot experiments,
			we have repeatedly observed throughout implementation 
			that processing times fluctuate only on the sub-second scale.
		}
		\label{tab:empirical_results}
	\end{center}
\end{table}

Experiment (I) and (II) were executed with the same  network configuration. 
The only difference was the separation of consensus and compute in the experiment (I),
while both were combined in the experiment (II). In our moderately simplified model,  
separating compute from consensus increased the throughput approximately by a \emph{factor} of 56.

Comparing experiment (II) and (III) illustrates the limited impact of increasing network resources compared. 
In terms of instructions per seconds, the network in the experiment (II) is 3.75 times more powerful%
\footnote{%
	 Let a slow node process $x$ instructions per second. 
	 Hence, under ideal resource utilization, the network in the experiment (III) can process $32x$ instructions. 
	 In contrast, the network in the experiment (III) processes $20x + 10 \cdot 5x + 2\cdot 25x = 120x$.
}
than in the experiment (III). However, the throughput of (II) increased only by $0.7\%$ compared to the  experiment (III). 
\medskip

\noindent
As the results show, separation of consensus and compute allows utilizing network resources more efficiently.  In a PoS network with combined consensus and compute, deterministic block finalization requires 
a super-majority of nodes to vote in favor of a candidate block to be finalized. 
For may BFT protocols, 
such as \cite{DworkLynchStockmeyer:PartialSynchronyConsensus:1988, Castro:PBFT:2002, buchman2016tendermint, HotStuff:2018}, 
finalization requires supporting votes with an accumulated fraction of stake $\mathcal{S} > \sfrac{2}{3}$. 
Though, some protocols have other limits, e.g.,\ $\mathcal{S} > 80\%$ for \cite{Martin:PAXOS:2006, David:Ripple:2014}. 
All of these protocols have in common that consensus nodes are obliged to execute the computation as a sub-task to verifying the proposed block.
Therefore, the time for finalizing a block is bound by the fastest $\mathcal{S}^\textnormal{th}$ percentile of nodes.
Less formally, the slowest nodes determine the throughput of the entire system. 
Consequently, running the network on stronger nodes leaves throughput unchanged as long as
the slowest $(1-\mathcal{S})$-fraction of nodes do not receive performance upgrades. 
In contrast, separation of consensus and compute significantly shifts computational load from the consensus nodes
to the fastest nodes in the network.


\section{Further Work\label{sec:further_work}}

A blockchain designed around the principles outlined in this paper would need to address additional problems, 
the most notable being the mechanism that verifies computation outputs.
On the one hand, splitting consensus and compute work boosts the throughput of \Architecture. 
On the other hand, special care has to be taken that the resulting states 
are correct as consensus nodes do not repeat the computation.

Furthermore, in \Architecture,
blocks no longer contain a hash commitment to the resulting state after computing the block.
Therefore, a node that receives data from a block state cannot verify the validity of the received data 
based on the information published in the  block. 
Nevertheless, a hash commitment for the result of a previous block 
can be published in a later block after passing verification. 
We will present the technical details of the \techterm{block verification} and 
commitment to the computation results (referred to as  \techterm{block sealing})
in the follow-up papers.
\medskip

\noindent%
The  presented simulations provide experimental evidence to support the theoretical work of this paper. 
While the theoretical results (section \ref{sec:TheoreticalAnalysis}) stand on their own without experimental validation, 
the experiments could be extended significantly. 
For example, we have not accounted for the extra steps required to verify computational states and commit them into the chain. 
Another aspect is the size of the consensus committee. 
It would be interesting to study the scaling of transaction throughput with different committees sizes of consensus and execution nodes. 
However, we have decided to prioritize implementing the \Architecture architecture over benchmarking a simplified model system. 
Throughput and other performance characteristics will be measured and published as soon as a full-fledged implementation is completed.

\section{Conclusions\label{sec:Conclusion}}

In this proof-of-concept work, we have demonstrated that a separation of  consensus and compute 
can lead to significantly increased resource utilization within PoS blockchain networks. 
For conventional round-based PoS networks, where one block is finalized before the subsequent block is proposed, 
the throughput is limited by a small fraction of the slowest nodes. 
In contrast, separation of consensus and compute significantly shifts computational load from the consensus nodes
to the fastest nodes in the network.  
We have shown in Theorem \ref{cor:delegation_of_work} that such separation of concern does not 
compromise the network's security. 
First experiments suggest that the throughput improvements enabled by such a separation of concerns are drastic. 
In a moderately simplified model, our simulations show a throughput increase by a factor of 56
compared to architectures with combined consensus and block computation.
\medskip

\noindent
One way to substantially increase the throughput of existing blockchains, such as Ethereum, could be to increase the gas limit.
However, this would accelerate the rate at which the state grows making it harder for new nodes to join the system. 
While in conventional proof-of-work blockchains the computational load to maintain and update the state is uniform across all (full) nodes,
the large majority of the computation resources are concentrated in a small fraction of mining nodes \cite{Miller:2015:BitcoinTopology}.

The \Architecture architecture  utilizes the  resource imbalance naturally occurring within a network ecosystem.
The few data-center-scale nodes with massive computational and bandwidth capacities can stake to become Execution Nodes 
to contribute their resources most efficiently. In contrast, Consensus Nodes do not store or maintain the state and,
therefore, can be run on off-the-shelf consumer hardware. 
With such separation of concerns, sharing a large state with new Execution Nodes joining the system should not pose a substantial challenge 
given the operational resources available to nodes with this role.

\clearpage
\addcontentsline{toc}{section}{Acknowledgments}
\section*{Acknowledgments}

We thank Dan Boneh for many insightful discussions, J.\, Ross Nicoll for contributions to an earlier draft,  
and 
Nick Johnson, 
Alex Bulkin, 
Karim Helmy, 
Teemu Paivinen, 
Travis Scher, 
Chris Dixon,
Jesse Walden,
Ali Yahya,
Ash Egan,
Casey Taylor,
Joey Krug,
Arianna Simpson,
as well as Lydia Hentschel
for reviews. 

\addcontentsline{toc}{section}{References}
\footnotesize
\bibliography{consensus_vs_compute}{}
\bibliographystyle{unsrt} 

\end{document}